\documentclass[%
    reprint,
    superscriptaddress,
    amsmath,amssymb, aps,
    prb,
    floatfix, longbibliography]{revtex4-2}

\usepackage{amsmath}
\usepackage{graphicx}
\usepackage{float}
\usepackage{dcolumn}
\usepackage{bm}
\usepackage{amssymb}
\usepackage[colorlinks=true,citecolor=blue]{hyperref}

\usepackage{physics}
\usepackage{array,multirow,makecell,longtable}
\usepackage[version=4]{mhchem}
\usepackage{siunitx}
\AtBeginDocument{\RenewCommandCopy\qty\SI}
\DeclareSIUnit\angstrom{\text{\AA}}

\providecommand{\peq}{\mathrel{\phantom{=}}\negmedspace{}}
\providecommand{\bk}[1]{\langle#1\rangle}
\providecommand{\degree}{^\circ}

\begin{document}

\title{Effects of Kitaev Interaction on Magnetic Orders and Anisotropy}

\author{Lianchuang Li}
\altaffiliation{Contributed equally to this work.}
\affiliation{
    Key Laboratory of Computational Physical Sciences (Ministry of Education),
    Institute of Computational Physical Sciences,
    State Key Laboratory of Surface Physics,
    and Department of Physics, Fudan University, Shanghai 200433, China.}

\author{Binhua Zhang}
\altaffiliation{Contributed equally to this work.}
\affiliation{
    Key Laboratory of Computational Physical Sciences (Ministry of Education),
    Institute of Computational Physical Sciences,
    State Key Laboratory of Surface Physics,
    and Department of Physics, Fudan University, Shanghai 200433, China.}

\author{Zefeng Chen}
\affiliation{
    Key Laboratory of Computational Physical Sciences (Ministry of Education),
    Institute of Computational Physical Sciences,
    State Key Laboratory of Surface Physics,
    and Department of Physics, Fudan University, Shanghai 200433, China.}

\author{Changsong Xu}
\email{csxu@fudan.edu.cn}
\affiliation{
    Key Laboratory of Computational Physical Sciences (Ministry of Education),
    Institute of Computational Physical Sciences,
    State Key Laboratory of Surface Physics,
    and Department of Physics, Fudan University, Shanghai 200433, China.}

\author{Hongjun Xiang}
\email{hxiang@fudan.edu.cn}
\affiliation{
    Key Laboratory of Computational Physical Sciences (Ministry of Education),
    Institute of Computational Physical Sciences,
    State Key Laboratory of Surface Physics,
    and Department of Physics, Fudan University, Shanghai 200433, China.}
\affiliation{Shanghai Branch, Hefei National Laboratory, Shanghai 201315, China}

\begin{abstract}
    We systematically investigate the effects of Kitaev interaction on magnetic
    orders and anisotropy in both triangular and honeycomb lattices. Our study
    highlights the critical role of the Kitaev interaction in modulating phase
    boundaries and predicting new phases, e.g., zigzag phase in triangular lattice and AABB phase in
    honeycomb lattice, which are absent with pure Heisenberg
    interactions. Moreover, we reveal the special state-dependent anisotropy
    of Kitaev interaction, and develop a general method that can determine the
    presence of Kitaev interaction in different magnets. It is found that the
    Kitaev interaction does not induce anisotropy in some magnetic orders such as
    ferromagnetic order, while can cause different anisotropy in other magnetic
    orders. Furthermore, we emphasize that the off-diagonal \(\Gamma\) interaction
    also contributes to anisotropy, competing with the Kitaev interaction to
    reorient spin arrangements. Our work establishes a framework for comprehensive
    understanding the impact of Kitaev interaction on ordered magnetism.
\end{abstract}

\maketitle

\section{Introduction}

The exactly solvable Kitaev model on the honeycomb lattice has recently
attracted significant attention due to its potential to host novel quantum
spin-liquid (QSL) states with Majorana fermion
excitations~\cite{kitaevAnyonsExactlySolved2006}. Jackeli and Khaliullin
suggested that the Kitaev interaction can be realized in Mott insulators with
edge-sharing octahedra, strong spin-orbit coupling (SOC), and electron
correlation~\cite{jackeliMottInsulatorsStrong2009}. Since then, there has been
a surge of theoretical and experimental studies on candidate Kitaev materials.
Established candidate materials include \ce{$A$2IrO3} ($A=\text{Na,
    Li}$)~\cite{caoEvolutionMagnetismSinglecrystal2013,manniEffectIsoelectronicDoping2014}
and
$\alpha$-\ce{RuCl3}~\cite{plumbRuCl3SpinorbitAssisted2014,kimKitaevMagnetismRuCl32015},
where the magnetic ions possess an effective spin $\tilde{S}=1/2$. Later,
cobalt compound with $3d^7$ configurations such as \ce{Na3Co2SbO6} were
proposed as candidate Kitaev materials with ferromagnetic Kitaev interaction,
which stems from the SOC of unquenched orbital angular momentum under small
crystal
field~\cite{liuPseudospinExchangeInteractions2018,liuKitaevSpinLiquid2020}.
Moreover, other $3d$ systems, such as \ce{CrI3},
\ce{CrGeTe3}~\cite{xuInterplayKitaevInteraction2018,xuPossibleKitaevQuantum2020}
and
\ce{NiI2}\cite{stavropoulosMicroscopicMechanismHigherSpin2019,amorosoSpontaneousSkyrmionicLattice2020a,liRealisticSpinModel2023},
were reported to exhibit significant Kitaev interaction via heavy ligand
mediated superexchange. These studies significantly expand the scope of the
so-called Kitaev system and pave the way for exploration into new realms of
inquiry.

However, these candidate materials exhibit ordered magnetism at low
temperatures rather than the expected QSL
state~\cite{liuLongrangeMagneticOrdering2011,searsMagneticOrderRuCl32015,kuindersmaMagneticStructuralInvestigations1981,huangLayerdependentFerromagnetismVan2017a,gongDiscoveryIntrinsicFerromagnetism2017},
sparking considerable research interest in the realistic effect of the Kitaev
interaction. For example, \ce{Na2IrO3} exhibits a collinear zigzag
antiferromagnetic (AFM) order with magnetic moments aligning along the
$44.3^\circ$ direction relative to the $a$ lattice vector, as determined by
diffuse magnetic X-ray scattering
experiments~\cite{hwanchunDirectEvidenceDominant2015}. In contrast,
$\alpha$-\ce{RuCl3} displays the same zigzag order, but its spins tend to
deviate by $35^\circ$ away from the $ab$ plane, as observed in X-ray
diffraction measurements~\cite{caoLowtemperatureCrystalMagnetic2016}. Moreover,
the low-temperature phase of bulk \ce{NiI2} is identified as a proper screw
helical state along $\bk{1\bar{1}0}$, with the normal of the helical plane
forming an angle of $55^\circ$ with the out-of-plane direction instead of
aligning along the in-plane propagation
direction~\cite{kuindersmaMagneticStructuralInvestigations1981,liRealisticSpinModel2023}.
Particularly, \ce{CrGeTe3} and \ce{CrI3} have ferromagnetic (FM) ground states
with a notable distinction: the former displays Heisenberg behavior, allowing
spins to orient freely in any
direction~\cite{gongDiscoveryIntrinsicFerromagnetism2017}, whereas the latter
behaves in an Ising manner along the out-of-plane
direction~\cite{huangLayerdependentFerromagnetismVan2017a}. These experimental
phenomena imply a cooperation between the Kitaev interaction $K$ and other
exchange interactions, such as isotropic exchange coupling
$J$~\cite{chaloupkaZigzagMagneticOrder2013,liRealisticSpinModel2023,xuInterplayKitaevInteraction2018}
and the off-diagonal exchange
$\Gamma$~\cite{kimKitaevMagnetismRuCl32015,wangTheoreticalInvestigationMagnetic2017},
thereby highlighting the significant impact of the Kitaev interaction on
magnetic orders and anisotropic spin orientation. However, to date, only a
handful of theoretical explorations of the Kitaev interaction in honeycomb
lattices (e.g., $J$-$K$ model and $J_1$-$J_2$-$J_3$-$K$ model in \ce{$A$2IrO3}
($A=\text{Na,
    Li}$)~\cite{chaloupkaZigzagMagneticOrder2013,kimchiKitaevHeisenbergJ2J3Model2011})
have been attempted. This calls for more comprehensive and qualitative
investigations of the Kitaev interaction in candidate triangular and honeycomb
systems, to establish general rules for its effect on magnetic behaviors.

In this work, we investigate the effect of Kitaev interaction on magnetic
orders within the $J_1\text{-}J_2\text{-}J_3\text{-}K$ model. We find that
introducing the Kitaev interaction can modulate the phase boundaries and
predict new phases absent in models with only Heisenberg interactions. In
addition, the Kitaev interaction exhibit special state-dependent anisotropy. It
can cause anisotropy in the magnetic orders lacking $C_3$ symmetry while does
not induce anisotropy in, for example, FM order. Furthermore, we reveal that
the off-diagonal $\Gamma$ term can modify the anisotropy relative to the pure
Kitaev interaction for various magnetic orders, which can account well for the
experimentally observed anisotropy of \ce{Na2IrO3} and $\alpha$-\ce{RuCl3}. Our
work provides useful insights for understanding the effects of the Kitaev
interaction on magnetic order and anisotropy.

\begin{figure}[tb]
    \centering
    \includegraphics[width=1\linewidth]{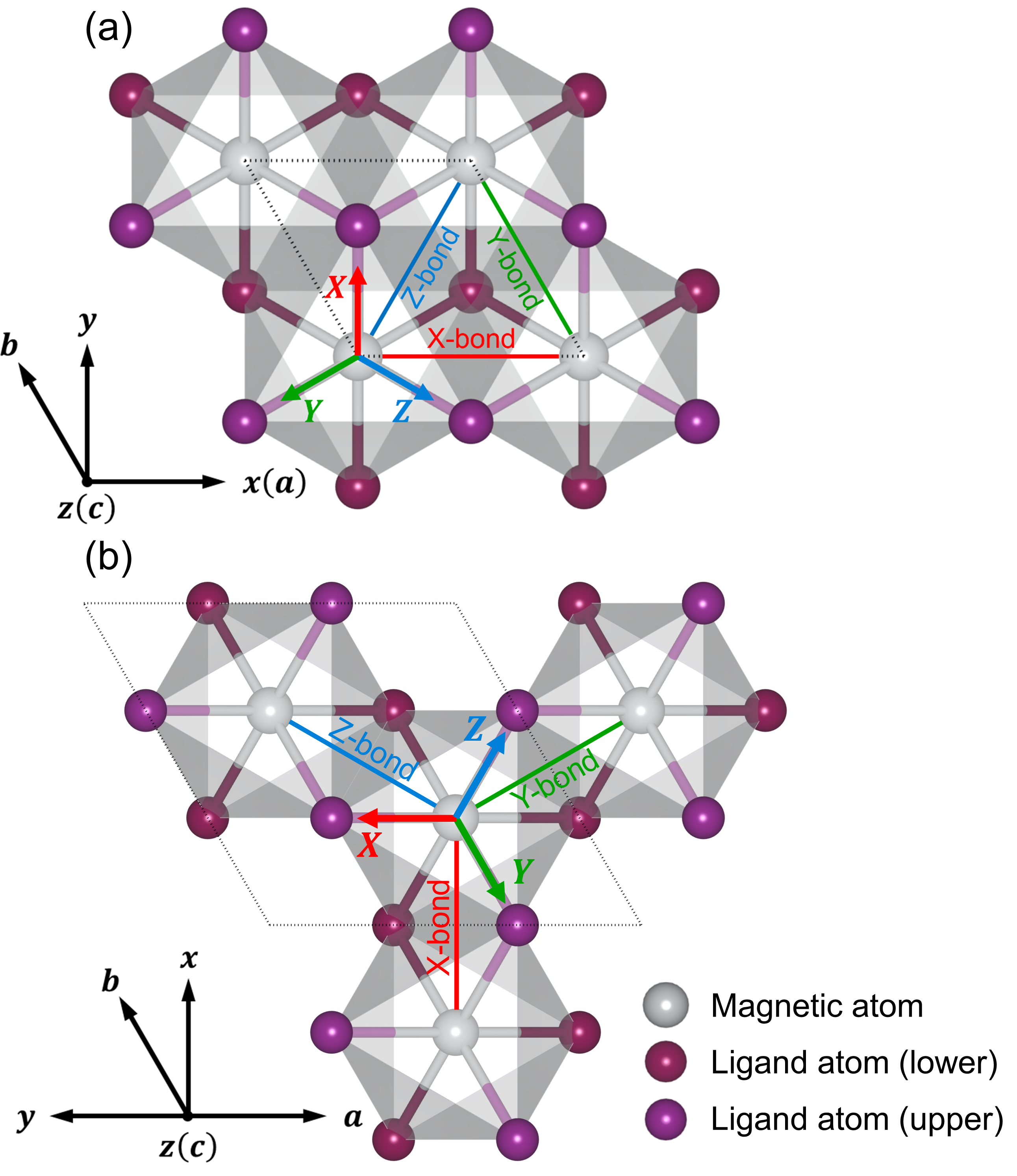}
    \caption{\label{fig:lattice}
        Schematization of the Kitaev basis of (a) triangular lattice and (b) honeycomb
        lattice. Dotted line marks the unit cell. The $\{XYZ\}$ basis of Kitaev is
        approximately along the direction of the bonds between magnetic atoms and upper
        ligand atoms. Kitaev basis and their corresponding bonds are indicated by red,
        green and blue colors. Note that the $X$ ($Y$, $Z$) axis of Kitaev basis is
        perpendicular to the X-bond (Y-bond, Z-bond, respectively).}
\end{figure}

\section{Effects of Kitaev interaction on magnetic orders}

The diverse magnetic orders observed in Kitaev candidate materials arise from
the interplay among different mechanisms, which are predominantly the
Heisenberg isotropic exchange interactions and the Kitaev interaction. To
delineate these possible magnetic behaviors, we define a spin Hamiltonian for
both triangular and honeycomb lattices as,
\begin{equation}
    \begin{split}
        H & =\sum_{\bk{i,j}_1}\left(J_1\vb{S}_i\cdot\vb{S}_j+KS_i^\gamma S_j^\gamma\right)                 \\
        &\peq+\sum_{\bk{i,j}_2}J_2\vb{S}_i\cdot\vb{S}_j+\sum_{\bk{i,j}_3}J_3\vb{S}_i\cdot\vb{S}_j,
    \end{split}
    \label{eq:J1J2J3K_model}
\end{equation}
where $\bk{i,j}$ denotes the pairs of interacting spins and $|\vb{S}|=1$ is assumed; $\gamma$ chooses its
value from three Kitaev basis \(XYZ\) and corresponds to X-bond, Y-bond and
Z-bond, respectively, as labeled in Fig.~\ref{fig:lattice}; $J$ and $K$
quantify the Heisenberg isotropy exchange interaction and the Kitaev
interaction, respectively, with positive values denoting AFM interactions and
negative ones representing FM interactions. Here the Heisenberg interactions up
to the third nearest neighbors (NN) are included in this model, as they are
usually important in honeycomb and triangular lattice.

\begin{figure*}
    \centering
    \includegraphics[width=1\linewidth]{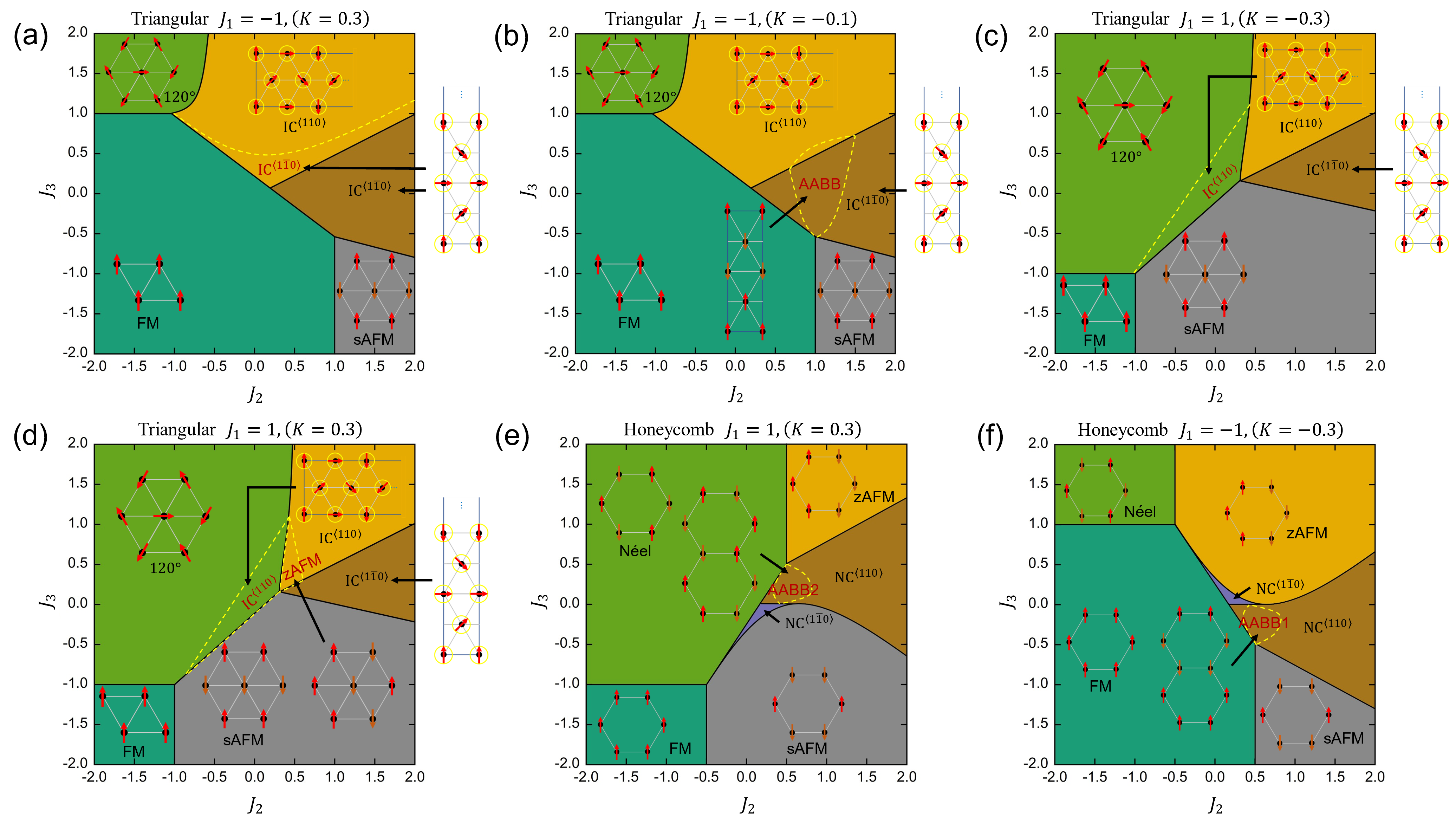}
    \caption{\label{fig:phase_diagrams}
        Phase diagrams of the $J_1\text{-}J_2\text{-}J_3(\text{-}K)$ model in
        triangular and honeycomb lattices. The colored areas in the phase diagrams show
        the ground states without Kitaev interaction. The yellow dashed lines indicate
        the most outstanding distinction of the ground states when introducing the
        Kitaev interaction.}
\end{figure*}

In order to determine the ground states, we first perform Monte Carlo (MC)
simulations over the $J_1\text{-}J_2\text{-}J_3\text{-}K$ model, as shown in
Eq.~\eqref{eq:J1J2J3K_model}. For the low energy states determined by MC,
conjugate gradient (CG) optimizations are further applied to obtain the
accurate spin structure and energy of different states. Both MC and CG methods
are implemented in the PASP software~\cite{louPASPPropertyAnalysis2021,SM}. For
simplicity, we fix the value of 1NN exchange interaction as $J_1=\pm1$, and
vary $J_2$ and $J_3$ among a range from $-2|J_1|$ to $2|J_1|$. We adopt
relatively weak values of $K$ (e.g. $K/|J_1|=\pm0.1$ or $K/|J_1|=\pm0.3$), so
as to prevent the dominant Kitaev from inducing QSL
state~\cite{kitaevAnyonsExactlySolved2006,hwanchunDirectEvidenceDominant2015},
which would fail with the classical MC approach. With these methods and
parameters, the obtained phase diagrams are displayed in
Fig.~\ref{fig:phase_diagrams}. Note that the the Luttinger-Tisza
method~\cite{freiserThermalVariationPitch1961,litvinLuttingerTiszaMethod1974}
is also used for determining the ground states of pure Heisenberg
$J_1\text{-}J_2\text{-}J_3$ models, and the results are consistent with our MC
and CG method.

We first investigate the phase diagram of pure Heisenberg
$J_1\text{-}J_2\text{-}J_3$ model in triangular lattice. In the case of the
triangular lattice, as depicted in Figs.~\ref{fig:phase_diagrams}(a-d), the FM
state prevails when $J_{2,3}\le-J_1$. For $J_2>-J_1$ and FM $J_3$, the system
tends to stabilize into the stripe AFM (sAFM) arrangement; while for $J_3>-J_1$
and FM $J_2$, the ground state adopts a $120\degree$ order (i.e., commensurate
helical state propagating along $\bk{110}$ with period $1.5a$), a hallmark of
frustration in the triangular lattice. Additionally, when both $J_2$ and $J_3$
are AFM, the system transitions to an incommensurate (IC) helical state. This
IC state propagates along $\bk{110}$ if $2J_2<J_3$, or along $\bk{1\bar{1}0}$
if $2J_2>J_3$.

The Kitaev interaction is then further considered, i.e., in the
$J_1\text{-}J_2\text{-}J_3\text{-}K$ model. As illustrated in
Fig.~\ref{fig:phase_diagrams}, the Kitaev effect on magnetic orders is evident
by the shifting of phase boundaries and the appearance of new phases, denoted
by the yellow dashed lines (minor shifts in other boundaries are neglected for
simplicity)\@. (i) Introducing AFM $K$ to the FM $J_1$ case
    [Fig.~\ref{fig:phase_diagrams}(a)] causes the expansion of the
$\text{IC}^{\bk{1\bar{1}0}}$ phase towards the $\text{IC}^{\bk{110}}$ phase,
even reaching across the entire boundary region between $\text{IC}^{\bk{110}}$
and FM with increasing $K$ [see Supplemental Materials (SM) Fig.~S1 for
        \(K=0.1\) case~\cite{SM}]. Note that such expansion can effectively account for
the experimentally observed $\text{IC}^{\bk{1\bar{1}0}}$ state in bulk
\ce{NiI2}, which will be predicted as an $\text{IC}^{\bk{110}}$ state under
pure Heisenberg interaction ($J_1=-4.976$ meV, $J_2=-0.155$ meV, $J_3=2.250$
meV)~\cite{liRealisticSpinModel2023}\@. (ii) Introducing FM $K$ to the FM $J_1$
case [Fig.~\ref{fig:phase_diagrams}(b)] leads to the emergence of a so-called
AABB AFM state with an up-up-down-down spin pattern along $\bk{1\bar{1}0}$
within the $\text{IC}^{\bk{1\bar{1}0}}$ region. This state rapidly replaces
most of the $\text{IC}^{\bk{1\bar{1}0}}$ phase with increasing $K$ (see SM
Fig.~S1 for \(K=-0.3\) case~\cite{SM})\@. (iii) When FM $K$ is introduced to
the AFM $J_1$ case [Fig.~\ref{fig:phase_diagrams}(c)], the $120\degree$ phase
near the border of $120\degree$-sAFM and $120\degree$-$\text{IC}^{\bk{110}}$
transitions into the $\text{IC}^{\bk{110}}$ phase with a propagation period
slightly smaller than $1.5a$\@. (iv) Introducing AFM $K$ to the AFM $J_1$ case
    [Fig.~\ref{fig:phase_diagrams}(d)] results in the enlargement of the
$\text{IC}^{\bk{110}}$ phase in the same region, particularly with a
propagation period slightly larger than $1.5a$. Meanwhile, the
$\text{IC}^{\bk{110}}$ phase near the border of
$120\degree$-$\text{IC}^{\bk{110}}$ transitions into a zigzag AFM (zAFM) state.

We now concentrate on the phase diagrams in the honeycomb lattice. As shown in
Figs.~\ref{fig:phase_diagrams}(e,f) (for more phase diagrams, see Fig.~S1 of
SM~\cite{SM}), the border between FM and other phases occurs at $J_3=-J_1$ and
$2J_2=-J_1$, which is different from $J_{2,3}=J_1$ in the triangular lattice.
This is understandable since the coordination number of the 2NN is twice than
that of 1NN in the honeycomb lattice. When $J_3>-J_1$ and $2J_2<-J_1$, the
N\'eel AFM state is stabilized, characterized by the $C_3$ symmetry akin to the
$120\degree$ order. For $2J_2>-J_1$ and FM $J_3$, the ground state turns out to
be the sAFM arrangement; while for $2J_2>J_1$ and AFM $J_3$, the system adopts
the zigzag AFM (zAFM) state. Note that two non-collinear (NC) states (see
Fig.~S2 of SM~\cite{SM} for detail) emerge between the sAFM and zAFM phases.
They propagate along $\bk{110}$ and $\bk{1\bar{1}0}$, respectively, and are
separated by $J_3=0$. Interestingly, introducing the $K/J_1>0$ Kitaev
interaction (see Fig.~S1 of SM~\cite{SM} for $K/J_1<0$ case) gives rise to new
collinear states, named AABB1 and AABB2, inside the
$\text{NC}^{\bk{1\bar{1}0}}$ region (the new collinear states region expands
very little with the increased $K/J_1>0$, see Fig.~S1 of SM~\cite{SM}). These
two new states exhibit the same up-up-down-down spin pattern on the zigzag
chain along $\bk{110}$, but behave as FM and AFM, respectively, along the
$\bk{1\bar{1}0}$ chain.

\begin{table*}[t]
    \caption{\label{tab:Kitaev_anisotropy}
    Anisotropy of the Kitaev interaction in triangular and honeycomb lattices. The
    orientation of Kitaev basis $\{XYZ\}$ is shown in the first column. The angle
    dependence of the Kitaev energy in a single unit cell is shown in the second
    column, where $c=\cos(\vb{k}\cdot\vb{R}_Y)$, $\vb{k}$ is the propagation vector
    and $\vb{R}_Y=\vb{b}$ is the lattice vector of a Y-bond.
    For example, with \(\text{IC}^{\bk{110}}\) order,
    \(\vb{k}=(\vb{a}+\vb{b})2\pi/\lambda a\) where $\lambda$ is the propagation
    period, yielding \(c=\cos{\pi a/\lambda}\) (see section~III.A of SM~\cite{SM} for more details).
    The spin configurations and their anisotropy preference under both FM and AFM Kitaev
    interaction are shown as well, where the red and orange arrows denote spins,
    blue arrows denote the normals of the helical plane, and blue circles indicate that
    the arrows (spins or norm vectors) can point to any directions in the plane
    without causing energy difference.}
    \begin{ruledtabular}
        \begin{tabular}{cccc}
            Lattice & Angle dependence of Kitaev energy                                        & FM $K<0$                                                                                        & AFM $K>0$ \\
            \hline
            \multirow{6}{*}[-35ex]{\makecell[c]{\includegraphics[width=0.09\textwidth]{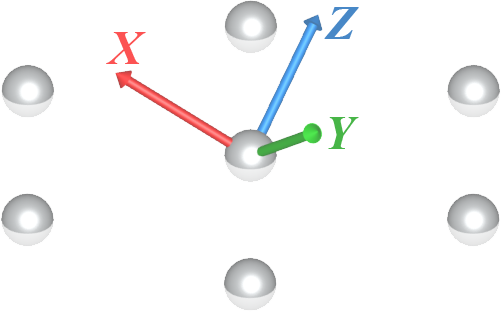}                                                                              \\Triangular}} & FM\@: no dependence                                                                            & isotropy                                                                                       & isotropy    \\
            \cline{2-4}
            ~       & sAFM\@: $2K\cos^2\alpha$                                                 & \makecell[c]{\includegraphics[width=0.09\textwidth]{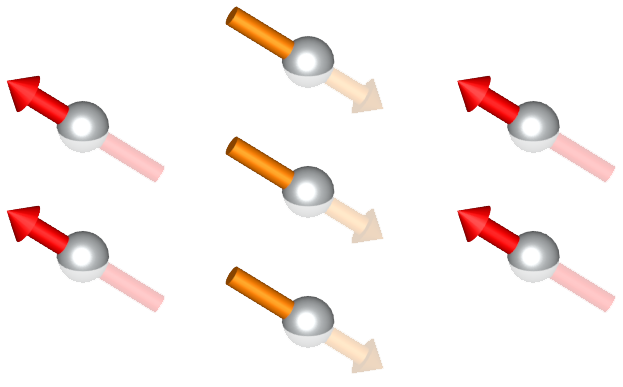}                      \\easy axis: $X$}                                                                      & \makecell[c]{\includegraphics[width=0.09\textwidth]{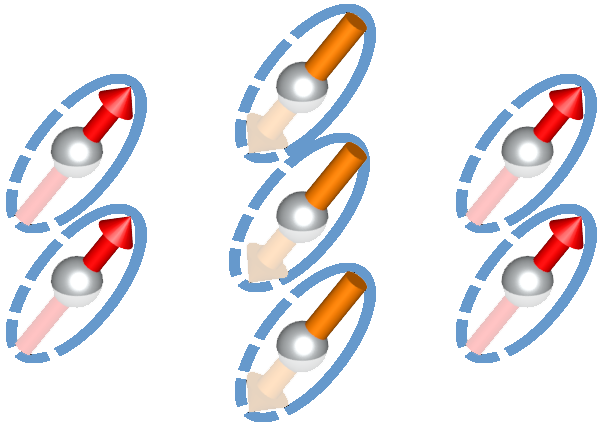}                \\easy plane: $Y$-$Z$} \\
            \cline{2-4}
            ~       & zAFM\@: $-K\cos^2\alpha$                                                 & \makecell[c]{\includegraphics[width=0.09\textwidth]{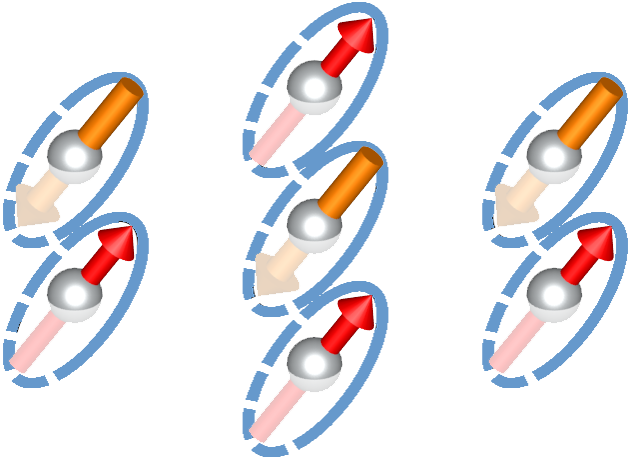}                     \\easy plane: $Y$-$Z$}                                                                           & \makecell[c]{\includegraphics[width=0.09\textwidth]{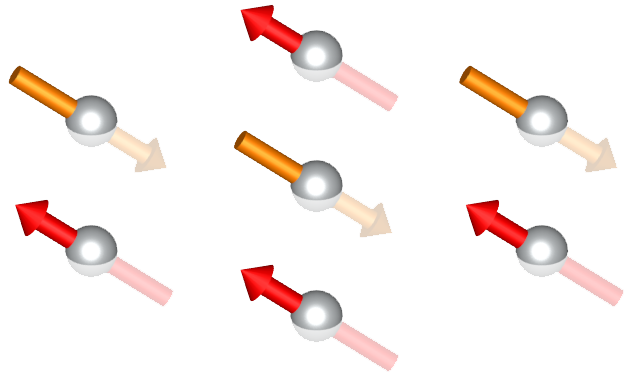}                \\easy axis: $X$}  \\
            \cline{2-4}
            ~       & AABB\@: $K\cos^2\alpha$                                                  & \makecell[c]{\includegraphics[width=0.09\textwidth]{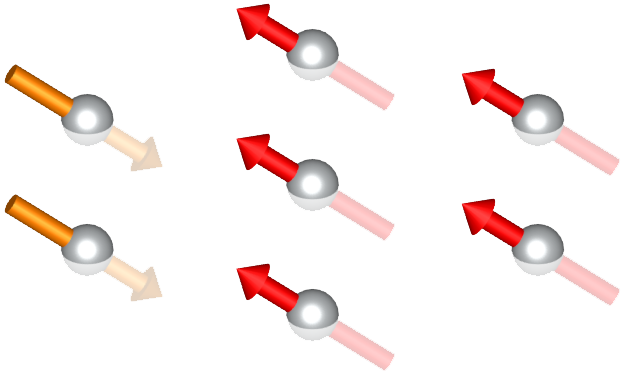}                      \\easy axis: $X$}                                                                             & \makecell[c]{\includegraphics[width=0.09\textwidth]{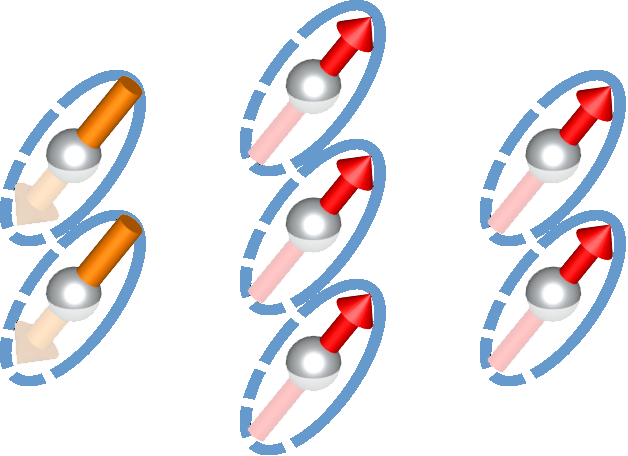}                \\easy plane: $Y$-$Z$} \\
            \cline{2-4}
            ~       & \multirow{3}{*}[-9ex]{$\text{IC}^{\bk{110}}$: $(2c^2-c-1)K\cos^2\alpha$} & \makecell[c]{\includegraphics[width=0.09\textwidth]{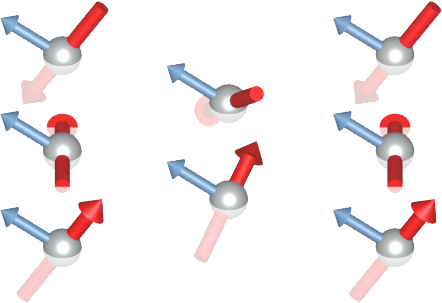}              \\$1.5a<\lambda<\infty$, normal along $X$}                                                       & \makecell[c]{\includegraphics[width=0.09\textwidth]{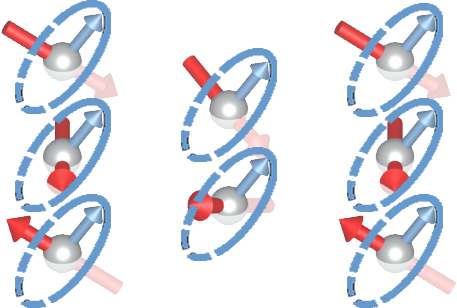}                                  \\$1.5a<\lambda<\infty$, normal in $Y$-$Z$ plane} \\
            ~       & ~                                                                        & \makecell[c]{\includegraphics[width=0.09\textwidth]{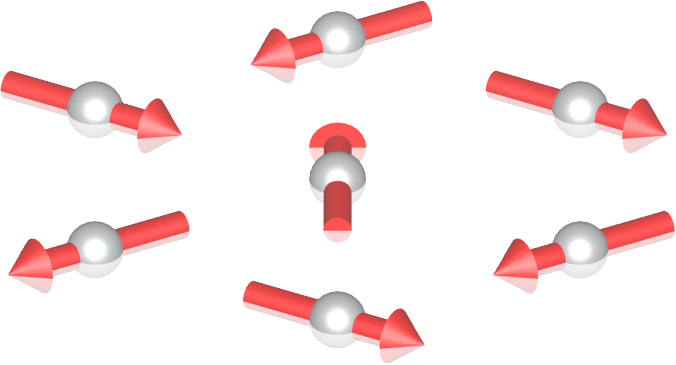}                         \\$\lambda=1.5a$, isotropy}                                                                        & \makecell[c]{\includegraphics[width=0.09\textwidth]{assets/spin_config/120_tri.png}                                  \\$\lambda=1.5a$, isotropy}        \\
            ~       & ~                                                                        & \makecell[c]{\includegraphics[width=0.09\textwidth]{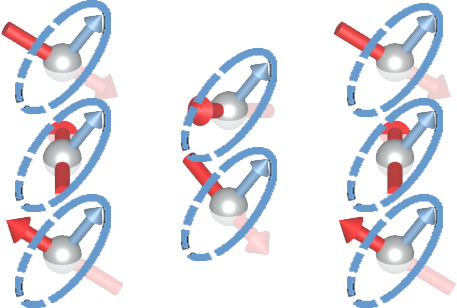}             \\$a\le\lambda<1.5a$, normal in $Y$-$Z$ plane}                                                               & \makecell[c]{\includegraphics[width=0.09\textwidth]{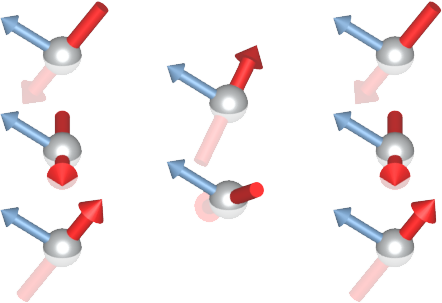}                                  \\$a\le\lambda<1.5a$,  normal along $X$} \\
            \cline{2-4}
            ~       & $\text{IC}^{\bk{1\bar{1}0}}$: $(1-c)K\cos^2\alpha$                       & \makecell[c]{\includegraphics[width=0.17\textwidth]{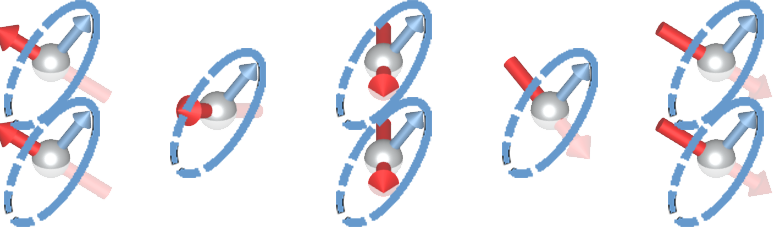}                   \\normal in $Y$-$Z$ plane}                                                                                    & \makecell[c]{\includegraphics[width=0.17\textwidth]{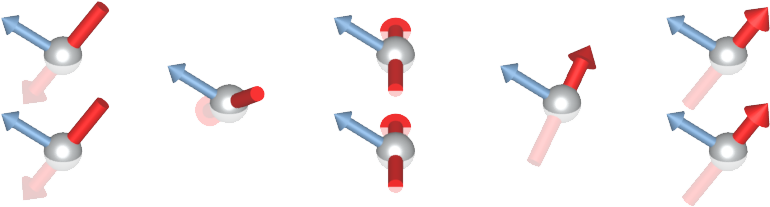}                                  \\normal along $X$}\\
            \hline
            \multirow{7}{*}[-20ex]{\makecell[c]{\includegraphics[width=0.09\textwidth]{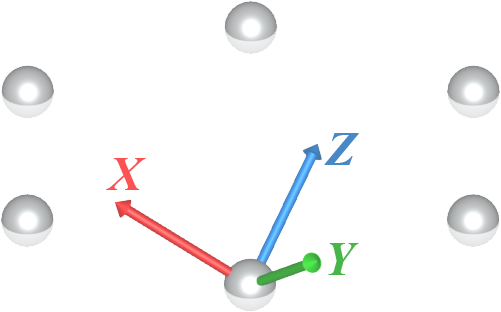}                                                                            \\Honeycomb}} & FM\@: no dependence                                                                            & isotropy                                                                                       & isotropy    \\
            \cline{2-4}
            ~       & sAFM\@: $2K\cos^2\alpha$                                                 & \makecell[c]{\includegraphics[width=0.09\textwidth]{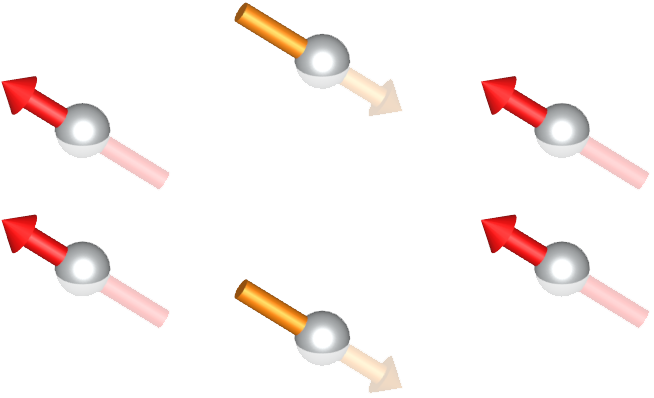}                    \\easy axis: $X$}                                                                      & \makecell[c]{\includegraphics[width=0.09\textwidth]{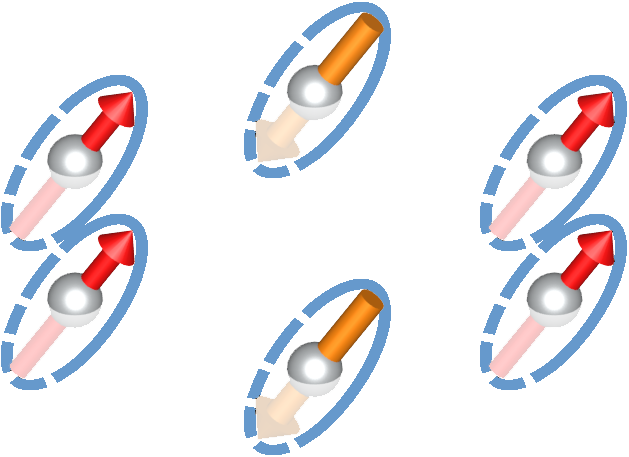}                \\easy plane: $Y$-$Z$} \\
            \cline{2-4}
            ~       & zAFM\@: $-2K\cos^2\alpha$                                                & \makecell[c]{\includegraphics[width=0.09\textwidth]{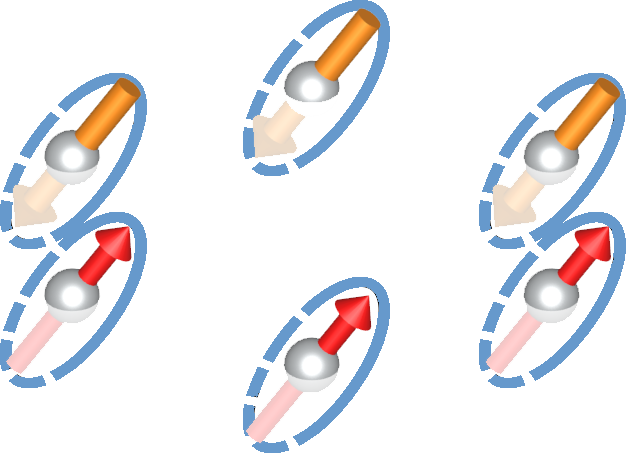}                   \\easy plane: $Y$-$Z$}                                                                           & \makecell[c]{\includegraphics[width=0.09\textwidth]{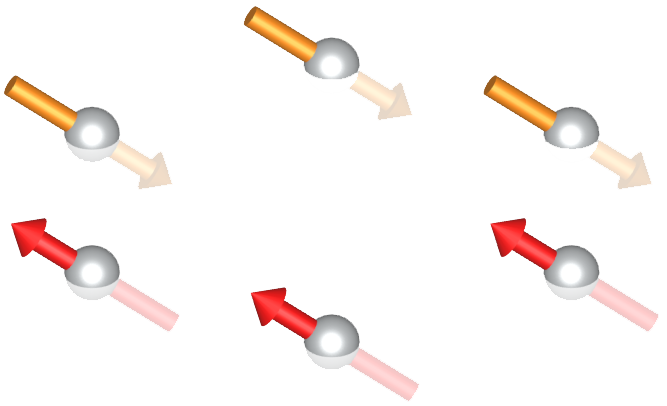}                \\easy axis: $X$}  \\
            \cline{2-4}
            ~       & AABB1: $K\cos^2\alpha$                                                   & \makecell[c]{\includegraphics[width=0.17\textwidth]{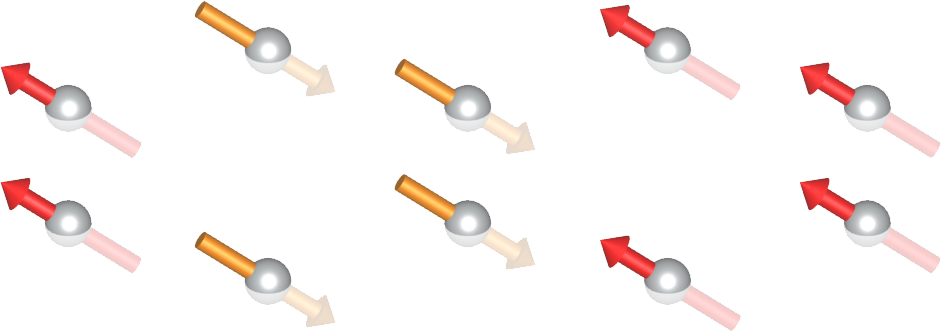}                   \\easy axis: $X$}                                                                             & \makecell[c]{\includegraphics[width=0.17\textwidth]{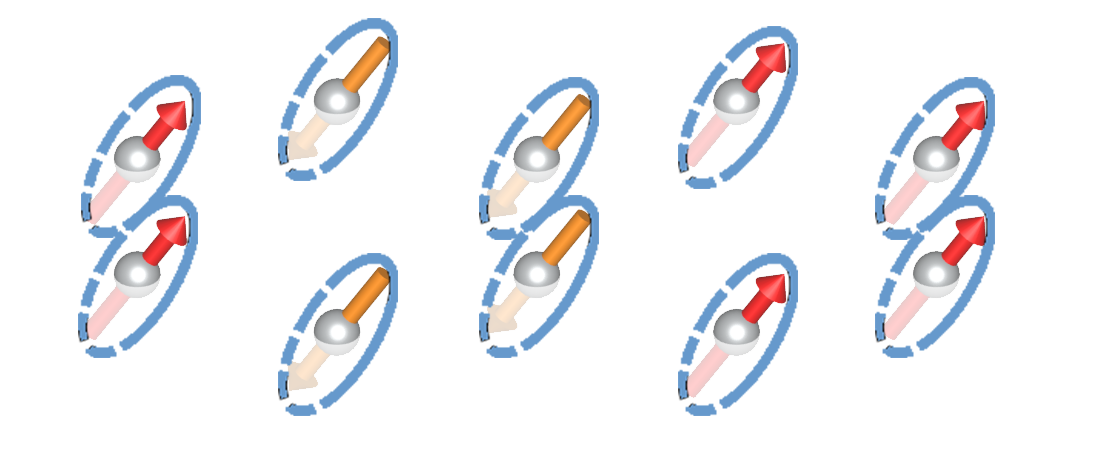}                \\easy plane: $Y$-$Z$} \\
            \cline{2-4}
            ~       & AABB2: $-K\cos^2\alpha$                                                  & \makecell[c]{\includegraphics[width=0.17\textwidth]{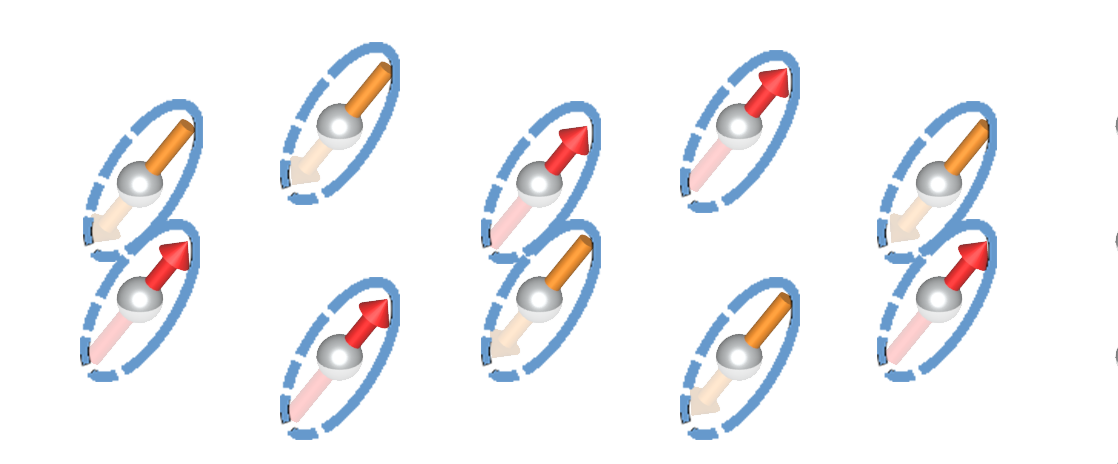}                  \\easy plane: $Y$-$Z$}                                                                             & \makecell[c]{\includegraphics[width=0.17\textwidth]{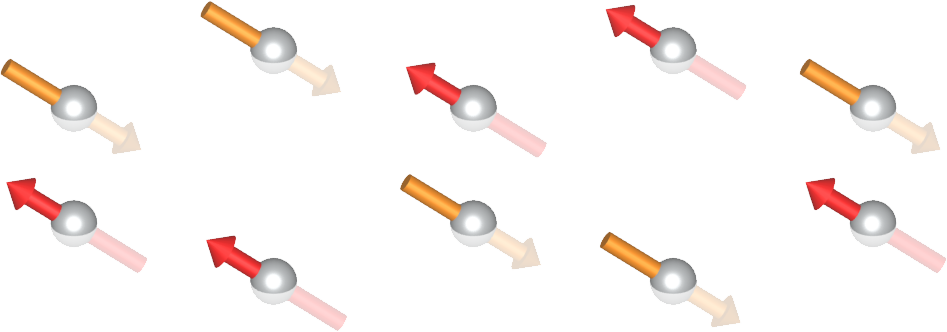}                \\easy axis: $X$} \\
            \cline{2-4}
            ~       & N\'eel: no dependence                                                    & \makecell[c]{\includegraphics[width=0.08\textwidth]{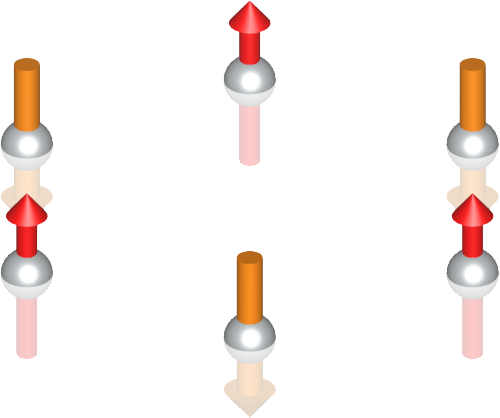}                      \\isotropy}                                                                                 & \makecell[c]{\includegraphics[width=0.08\textwidth]{assets/spin_config/Neel_honey.png}                \\isotropy}
        \end{tabular}
    \end{ruledtabular}
\end{table*}

\section{Effects of Kitaev interaction on anisotropy}

We notice that the aforementioned magnetic orders exhibit novel anisotropy when
Kitaev interaction comes into play. Such anisotropy arising from Kitaev
interaction, not only changes when Kitaev interaction changes sign, but also
varies when the magnetic state becomes different. It is thus a rare
state-dependent anisotropy, based on which a method is recently proposed to
predict the presence of Kitaev interactions in
Ref.~\cite{chenStrengthKitaevInteraction}. Here we perform systematical
analysis on the state-dependent Kitaev anisotropy. Taking the zAFM state in
honeycomb lattice as an example, the energy contribution from Kitaev
interaction of
a unit cell
can be rewritten as a function of spin orientation,
\begin{equation}
    \begin{split}
        E_K & =\sum_{\bk{i,j}_1}KS_i^{\gamma}S_j^{\gamma} \\
        & =-KS^{X}S^{X}+KS^{Y}S^{Y}+KS^{Z}S^{Z}       \\
        & =KS^2-2K{(S^X)}^2                           \\
        & =KS^2-2S^2K\cos^2\alpha\\
        & \varpropto -2K\cos^2\alpha\text{.}
    \end{split}
    \label{eq:E_K_honey_zAFM}
\end{equation}
where $\alpha$ is the angle between the spin $\vb{S}$ and $X$ axis of Kitaev
basis. Note that the present exampled zAFM order exhibits AFM link along X-bond while FM link along
Y-bond and Z-bond.
After minimizing the Eq.~\eqref{eq:E_K_honey_zAFM}, we can obtain: when
$K<0$, it results in $\alpha=90\degree$, indicating the spins favor lying in
the $Y$-$Z$ plane (easy plane); whereas $K>0$, it leads to $\alpha=0\degree$,
corresponding to the $X$ direction as the easy axis.

Table.~\ref{tab:Kitaev_anisotropy} summaries the anisotropy of spin
orientations in different magnetic orders (details are given in section~III.A
of SM~\cite{SM}). The table shows that (i) the anisotropy of the Kitaev
interaction is indeed state-dependent, where different magnetic ordered states
may exhibit different anisotropies, even for the same \(K\); (ii) within a same
order, anisotropy changes when the Kitaev interaction changes its sign. For
example, when $K>0$, zAFM state in honeycomb lattice has an easy axis along the
Kitaev basis \(X\), while for sAFM state in honeycomb lattice, the \(X\) basis
becomes a hard axis. And if the Kitaev interaction changes the sign, the easy
axis will become hard axis, and vice versa. Noteworthily, most states exhibit
the Kitaev-induced anisotropy with Kitaev basis being easy/hard axis, except
for the $\text{IC}^{\bk{110}}$ state with period $\lambda=1.5a$ (i.e.,
$120\degree$ state) in triangular lattice, N\'eel state in honeycomb lattice,
and FM state in both triangular and honeycomb lattices. This is understandable
since all of these excluded states possess the $C_3$ symmetry, resulting in the
same Kitaev energy for pairs connected by X-, Y-, and Z-bonds, thereby leading
to the isotropic Kitaev effect in these states.

\begin{figure}
    \centering
    \includegraphics[width=1\linewidth]{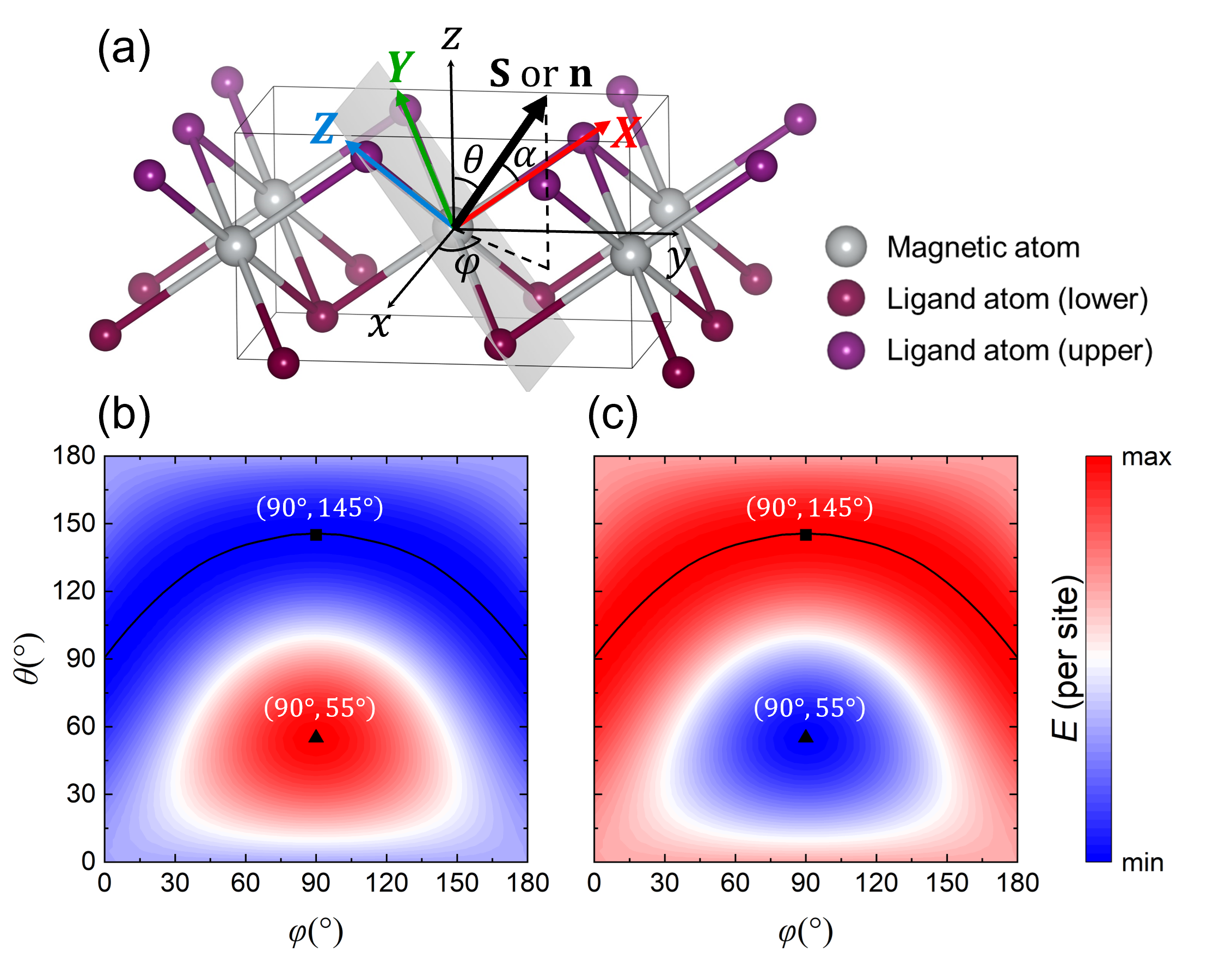}
    \caption{\label{fig:energy_diagrams}
        Magnetic anisotropy energy (MAE) of pure Kitaev interaction applied to
        different spin configurations\@. (a) Schematizes the definitions of
        $\theta,\varphi,\alpha$ and Kitaev basis. $z$ axis is defined as the out-of-plane
        direction and $y$ axis is along the direction of in-plane projection of $X$
        axis\@. Panel (b) and (c) are the MAE diagrams ignoring specific energy values.
        For collinear states, the MAE is represented by the direction of spin vector
        $\vb{S}$, and for helical states, by the direction of the normal vector of
        helical plane $\vb{n}$\@. Panel (b) shows the MAE of the spin configurations
        preferring ``$Y$-$Z$ plane'' and panel (c) shows the MAE of the spin
        configurations preferring ``$X$'' axis. The triangular mark and black curve
        locate the extreme points.}
\end{figure}

The energy expressions in Table.~\ref{tab:Kitaev_anisotropy} are based on a
unit cell. We now exam the energy of the entire supercell, which can
accommodate a complete period of spin structures, to further validate the
aforementioned anisotropy. Simulations is conducted for of the magnetic
anisotropy energy (MAE) from pure Kitaev interaction across various magnetic
orders and propagation periods in Table.~\ref{tab:Kitaev_anisotropy} (see
section~I.B of SM~\cite{SM} for detail). It is found that, though there are
various spin orders together with positive/negative Kitaev interactions, the
energy distribution can be summarized by two kinds of MAE diagrams, with easy
plane and easy axis respectively. The summarized all possible energy
distributions (excluding trivial isotropy cases) are plotted in
Fig.~\ref{fig:energy_diagrams}. It is clear to see that there are
minimum/maximum value aligned with Kitaev basis ($\varphi=90^\circ$,
$\theta=55^\circ$), indicating it as the easy/hard axis, thus confirming
results in Table.~\ref{tab:Kitaev_anisotropy} once again.

Based on the conclusions of Table.~\ref{tab:Kitaev_anisotropy}, the anisotropy
of several Kitaev candidates can be understood. For the FM states in
\ce{CrGeTe3} and \ce{CrI3}, the Kitaev interaction does not influence
anisotropy in the FM configuration. Therefore, the differing Heisenberg or
Ising behaviors observed in the FM states of \ce{CrGeTe3} and \ce{CrI3} should
be attributed to other interactions, such as the off-diagonal $\Gamma$
interaction and single-ion anisotropy (SIA) (for more details, see
section~IV(a) of SM~\cite{SM}), rather than the pure Kitaev interaction. In
contrast, Kitaev interaction plays an important role in the anisotropy of
\ce{NiI2} as the $55\degree$ canting of the normal of the rotation plane in the
helical ground state can be accurately predicted by an AFM Kitaev interaction,
which was confirmed by density functional theory (DFT)
calculations~\cite{liRealisticSpinModel2023}.

\begin{figure*}[t]
    \centering
    \includegraphics[width=1\linewidth]{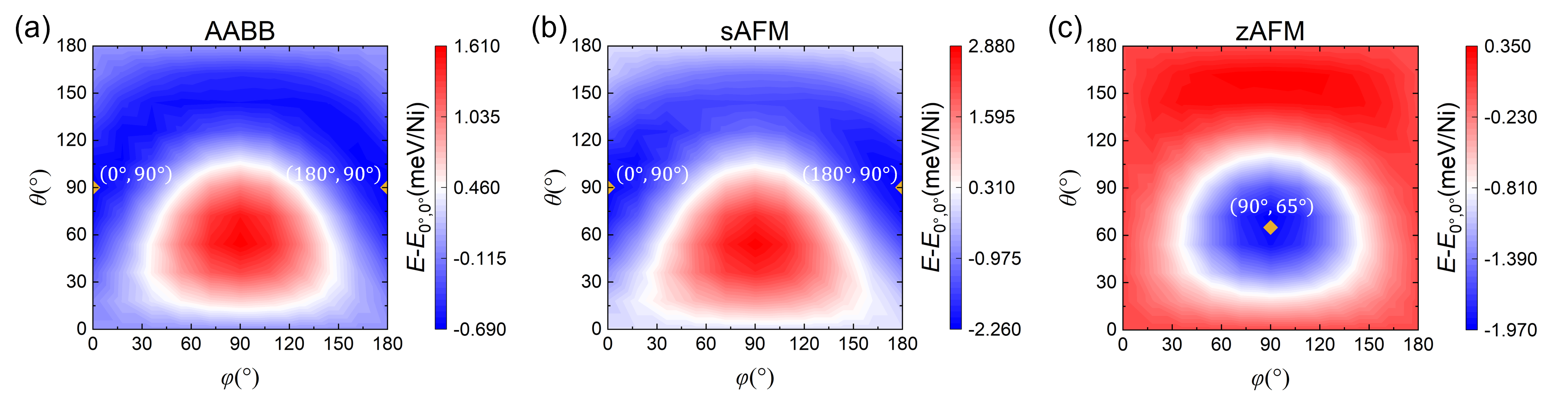}
    \caption{\label{fig:NiI2_energy_diagrams}
        Magnetic anisotropy energy of monolayer \ce{NiI2} with spin configurations of
        (a) AABB, (b) sAFM and (c) zAFM\@. Definition of $\varphi$ and $\theta$ are shown
        in Fig.~\ref{fig:energy_diagrams}(a). Relative energy with respect to energy at $(\varphi, \theta)$
        $(0^\circ,0^\circ)$ (namely, $z$ direction) is plotted. The diamond marks
        locate the minimum of energy.}
\end{figure*}

\subsection{EFFECTS OF $\Gamma$ INTERACTION ON ANISOTROPY}

Despite the success of the Kitaev interaction in explaining the anisotropy in
\ce{NiI2}, the spin orientations of zigzag AFM in \ce{Na2IrO3} and
$\alpha$-\ce{RuCl3} cannot be fully reproduced by the Kitaev interaction alone,
implying the effects of other anisotropy. Besides the pure Kitaev interaction,
the off-diagonal exchange interaction $\Gamma$ can also modulate the spin
orientation, thereby inducing additional anisotropy. By extending Kitaev model
$KS_i^{\gamma}S_j^{\gamma}$ to incorporate such anisotropy mechanisms, the
Kitaev related Hamiltonian becomes~\cite{winterModelsMaterialsGeneralized2017}
\begin{equation}
    \begin{split}
        H & =\sum_{\bk{i,j}_1}\left\{
        KS_i^\gamma S_j^\gamma
        +\Gamma\left(S_i^\alpha S_j^\beta+S_i^\beta S_j^\alpha\right)\right.\\
        &\left.\peq+\Gamma^\prime\left(
        S_i^\gamma S_j^\alpha+S_i^\gamma S_j^\beta
        +S_i^\alpha S_j^\gamma+S_i^\beta S_j^\gamma
        \right)\right\}\\
        &=\sum_{\bk{i,j}_1}\vb{S}_i^T\mathcal{K}_\gamma\vb{S}_j,
    \end{split}
\end{equation}
where $\{\alpha\beta\gamma\}=\{YZX\},\{ZXY\}$ and $\{XYZ\}$ for X-, Y- and
Z-bonds, respectively. $\mathcal{K}_\gamma=\mathcal{K}_{X,Y,Z}$ refers to the
Kitaev interaction matrices in the Kitaev basis shown in
Fig.~\ref{fig:lattice}, which have the following forms
\begin{equation}
    \begin{pmatrix}
        K             & \Gamma^\prime & \Gamma^\prime \\
        \Gamma^\prime & 0             & \Gamma        \\
        \Gamma^\prime & \Gamma        & 0
    \end{pmatrix},
    \begin{pmatrix}
        0             & \Gamma^\prime & \Gamma        \\
        \Gamma^\prime & K             & \Gamma^\prime \\
        \Gamma        & \Gamma^\prime & 0
    \end{pmatrix},
    \begin{pmatrix}
        0             & \Gamma        & \Gamma^\prime \\
        \Gamma        & 0             & \Gamma^\prime \\
        \Gamma^\prime & \Gamma^\prime & K
    \end{pmatrix}\text{.}
\end{equation}
The $\Gamma$ term, stemming from the orbital coupling, can persist in cubic
octahedra~\cite{katukuriKitaevInteractionsMoments2014,rauGenericSpinModel2014}.
While the $\Gamma^\prime$ term is expected to arise from trigonal distortion of
octahedra~\cite{chaloupkaHiddenSymmetriesExtended2015}.
Note that the idealized undistorted crystal structures were adopted in our
analysis and calculations, thus the following discussion will mainly focus on
the anisotropy effect of \(\Gamma\) term.

The $\Gamma$ term can modify the anisotropy relative to the pure Kitaev
interaction for various magnetic orders (see section~IV of SM for more
details~\cite{SM}). Considering the zAFM order
propagating along X-bond in honeycomb lattice as illustrated in
Table.~\ref{tab:Kitaev_anisotropy}, when $K<0$ and $\Gamma<0$ (both FM), the
easy axis will be fixed at $x$ direction of the global $\{xyz\}$ basis, while
in other cases the direction of easy axis will be tilted by $\Gamma$. In such
scenario, the angle $\theta$ of easy axis can be evaluated by
(see section~IV(e) of SM~\cite{SM} for the deduction)
\begin{equation}
    \theta=90\degree-\arctan{
        \frac{7\Gamma+2K-3\sqrt{9\Gamma^2-4K\Gamma+4K^2}}{4\sqrt{2}(K-\Gamma)}
    }
    \label{eq:theta_K_Gamma}
\end{equation}
with \(\phi=90\degree\). For the FM $K<0$ and AFM $\Gamma>0$, as $\Gamma$
increases from zero, the spin orientation will be tilted from the angular
bisector direction of $Y,Z$ axis (i.e., $\varphi=90\degree$ and
$\theta=145\degree$) with the angle $\theta$ decreasing. In the case of AFM
$K>0$, an increasing the magnitude of FM $\Gamma<0$ will slightly drive the
spin away from $X$ direction (i.e., $\varphi=90\degree$ and $\theta=55\degree$)
by decreasing the angle $\theta$; while increasing AFM $\Gamma>0$ results in
the opposite effect, i.e., causing angle $\theta$ to increase.

According to Eq.~\eqref{eq:theta_K_Gamma}, the spin canting angles of
$44.3\degree$ and $35\degree$ away from the $ab$ plane in \ce{Na2IrO3} and
$\alpha$-\ce{RuCl3}, respectively, can be reproduced. For \ce{Na2IrO3}, the
angle of $44.3\degree$ (i.e., $\theta=134.3\degree$) requires a FM $K<0$ and
AFM $\Gamma>0$ with $K/\Gamma=-3.21$ in absent of other anisotropy terms. For
$\alpha$-\ce{RuCl3}, adopting the results of $K=-6.8$ meV and $\Gamma=9.5$ meV
given by the fitting of inelastic neutron scattering
measurements~\cite{ranSpinWaveExcitationsEvidencing2017}, the angle will be
determined to about $30.1\degree$, very close to the experiment result
$35\degree$.

To exam the anisotropic effects of the Kitaev and $\Gamma$ interactions, we now
perform DFT calculations on monolayer NiI$_2$, which is proposed to be a Kitaev
dominant system.
The DFT calculations on a monolayer \ce{NiI2}, adopting AABB, sAFM and zAFM
configurations, are performed, using
PBE+U~\cite{perdewGeneralizedGradientApproximation1996}
\nocite{blochlProjectorAugmentedwaveMethod1994,kresseEfficientIterativeSchemes1996,haceneAcceleratingVASPElectronic2012,hutchinsonVASPGPUApplication2012},
and the results are shown in Fig.~\ref{fig:NiI2_energy_diagrams}. These
diagrams resemble Fig.~\ref{fig:energy_diagrams}(b,c) very much, implying the
dominance of Kitaev anisotropy.
For AABB and zAFM state, the global magnetic anisotropy energies are 2.3 meV/Ni
and 2.32 meV/Ni, respectively, while for the sAFM state, the anisotropy energy
is 5.14 meV/Ni, nearly twice that of the AABB and zAFM states. This observation
is consistent with the Kitaev energy values listed in
Table.~\ref{tab:Kitaev_anisotropy}\@.
According to the method of fitting anisotropic interactions proposed by
Ref.~\cite{chenStrengthKitaevInteraction}, an approximation of the Kitaev
interaction fitted by data in Fig.~\ref{fig:NiI2_energy_diagrams} is
\(K=2.27\)meV.
The direction of easy axis in zAFM is found to locate at $(\varphi,
    \theta)=(90\degree,65\degree)$, close to the prediction of
$(90\degree,55\degree)$ from Kitaev interaction, with a deviation of
$\Delta\theta=10\degree$. In contrast, the easy axis of AABB and sAFM are both
determined at $(0\degree,90\degree)$ or $(180\degree,90\degree)$ which is
exactly $\pm x$ direction, while the energy values at the plane perpendicular
to $(90\degree,55\degree)$ are basically constant. Such actual anisotropy
matches well with the easy plane anisotropy of the Kitaev interaction, only
with the breaking of degeneracy into $x$ direction. In fact, the deviation in
anisotropy energy and easy axis, as well as the breaking of degeneracy in the
easy plane, can be attributed to a weak in-plane SIA
term~\cite{liRealisticSpinModel2023}, which slightly modulates the MAE
distribution but still retains the characteristic features of Kitaev
anisotropy. Therefore, these results not only verify our theory presented in
this work, but also support the presence of Kitaev interaction in
NiI$_2$~\cite{stavropoulosMicroscopicMechanismHigherSpin2019,amorosoSpontaneousSkyrmionicLattice2020a,liRealisticSpinModel2023}.

\section{Conclusions}
To conclude, we have studied the effects of Kitaev interaction on the magnetic
order of Heisenberg-Kitaev $J_1\text{-}J_2\text{-}J_3\text{-}K$ model in both
triangular and honeycomb lattice by utilizing Monte Carlo simulations and
conjugate gradient optimizations. The results of $J_2\text{-}J_3$ phase
diagrams suggest that weak Kitaev interaction will enlarge the incommensurate
state region and lead to new phases such as zigzag and AABB phase in different
cases. Our analysis and calculations demonstrate the state-dependent anisotropy
of the Kitaev interaction. With a pure Kitaev interaction, all the spin
configurations presenting in the phase diagrams, except the FM, $120\degree$
and N\'eel states, which possess $C_3$ symmetry, will choose the Kitaev basis
to be an easy/hard axis. This excludes the effects of Kitaev interaction on the
anisotropy of FM state in \ce{CrGeTe3} and \ce{CrI3}, but emphasizes the key
role of Kitaev interaction in the canted proper screw state of \ce{NiI2}. For
the off-diagonal $\Gamma$ interaction, it can modify the anisotropy relative to
the pure Kitaev interaction, and slightly tilt the easy axis of zigzag AFM
state in honeycomb lattice, which can well explain the experimentally observed
anisotropy of $\alpha$-\ce{RuCl3} and \ce{Na2IrO3}. Our work thus provides
valuable insights into the effects of the Kitaev interaction on magnetic order
and anisotropy.

\begin{acknowledgments}
    We acknowledge financial support from the National Key R\&D Program of China
    (No. 2022YFA1402901), NSFC (Grants No. 11825403, No. 11991061, No. 12188101,
    No. 12174060, and No. 12274082), the Guangdong Major Project of the Basic and
    Applied basic Research (Future functional materials under extreme
    conditions-C2021B0301030005), Shanghai Pilot Program for Basic Research-FuDan
    University 21TQ1400100 (23TQ017), and Shanghai Science and Technology Program
    (23JC1400900). C. X. also acknowledges support from the Shanghai Science and
    Technology Committee (Grant No. 23ZR1406600). B. Z. also acknowledges the
    support from the China Postdoctoral Science Foundation (Grant No. 2022M720816).
\end{acknowledgments}

\bibliography{main}

\end{document}